\documentclass{article}
\usepackage{spconf,amsmath,graphicx}
\usepackage{amsfonts}
\usepackage{hyperref}
\usepackage{setspace}

\usepackage{color}
\usepackage{xcolor}
\usepackage{threeparttable}
\usepackage{multirow}
\usepackage{multicol}
\usepackage{booktabs}

\title{LCB-net: Long-Context Biasing for Audio-Visual Speech Recognition}
\name{\begin{tabular}{c}Fan Yu, Haoxu Wang, Xian Shi, Shiliang Zhang\end{tabular}}
\address{Speech Lab of DAMO Academy, Alibaba Group, China 
        }

\begin{document}
%
\maketitle
\begin{abstract}

The growing prevalence of online conferences and courses presents a new challenge in improving automatic speech recognition (ASR) with enriched textual information from video slides.
In contrast to rare phrase lists, the slides within videos are synchronized in real-time with the speech, enabling the extraction of long contextual bias.
Therefore, we propose a novel long-context biasing network (LCB-net) for audio-visual speech recognition (AVSR) to leverage the long-context information available in videos effectively.
Specifically, we adopt a bi-encoder architecture to simultaneously model audio and long-context biasing.
Besides, we also propose a biasing prediction module that utilizes binary cross entropy (BCE) loss to explicitly determine biased phrases in the long-context biasing.
Furthermore, we introduce a dynamic contextual phrases simulation to enhance the generalization and robustness of our LCB-net.
Experiments on the SlideSpeech, a large-scale audio-visual corpus enriched with slides, reveal that our proposed LCB-net outperforms general ASR model by 9.4\%/9.1\%/10.9\% relative WER/U-WER/B-WER reduction on test set, which enjoys high unbiased and biased performance.
Moreover, we also evaluate our model on LibriSpeech corpus, leading to 23.8\%/19.2\%/35.4\% relative WER/U-WER/B-WER reduction over the ASR model.

\end{abstract}
\begin{keywords}
Multi-Modal, Audio-Visual Speech Recognition, Biasing Prediction, Contextual Adaptation
\end{keywords}
\vspace{-8pt}
\section{Introduction}
\label{sec:intro}
\vspace{-4pt}
In recent years, with the advances in deep learning, end-to-end (E2E) models have achieved great success in automatic speech recognition (ASR), including connectionist temporal classification (CTC)~\cite{graves2006connectionist}, listen-attend-and spell (LAS)~\cite{chan2016listen}, Transducer~\cite{graves2013speech} and Transformer~\cite{DBLP:journals/corr/VaswaniSPUJGKP17}.
ASR is widely applied to streaming media (e.g. online conferences and courses) which consists of both an audio and a visual stream.
However, ASR often struggles to accurately transcribe rare phrases, such as personal names, technical terms, place names, entity names, and other phrases~\cite{wudual}. 
Fortunately, many researchers have demonstrated that contextual information is crucial to accurately predict rare phrases~\cite{cpp2023Huang}. 
Especially, within these applications, the slides featured in the visual stream can provide valuable textual information for improving ASR, particularly when the audio includes rare phrases.
Consequently, contextual biasing models, which incorporate contextual knowledge into ASR, have been extensively developed~\cite{cpp2023Huang,williams2018contextual,chen2019end,deepcontext}.

Numerous research studies are dedicated to improving the performance of ASR on contextual phrases by incorporating contextual and biased information.
Shallow fusion~\cite{williams2018contextual,chen2019end,zhao2019shallow} is a representative method to bias the decoding process towards contextual phrases by integrating an independently trained language model (LM).
The LM adjusts the posterior probabilities of contextual phrases.
However, shallow fusion requires retraining an LM for different contextual biasing lists and determining the optimal fusion weight for various scenarios, which consumes considerable time and computational resources.
To improve the flexibility of biasing methods, joint optimization of ASR and contextual module has been proposed~\cite{cpp2023Huang,deepcontext,contextrnnt,han2021cif,chen2019joint,han2022improving}.
Within the joint learning framework, E2E contextual ASR model can perform biasing edits in latent features, enabling rapid adaptation to different contextual biasing lists.
Contextual listen, attend and spell (CLAS)~\cite{deepcontext} and context-ware Transformer Transducer (CATT)~\cite{contextrnnt} introduce attention mechanisms to capture the relationship between ASR features and contextual biasing embeddings.
Although CLAS and CATT efficiently model the contextual information and achieve significant improvements comparison to shallow fusion, they lack explicit modeling for biasing prediction tasks.
Recently, Huang et al.~\cite{cpp2023Huang} and Han et al.~\cite{han2021cif} propose contextual phrase prediction (CPP) and continuous integrate-and-fire based collaborative decoding (CIF ColDec) network, respectively, to explicitly predict bias. These approaches enforce ASR encoder to pay more attention to contextual bias.

\begin{figure*}[!t]
	\centering
        \vspace{-7pt}
	\includegraphics[scale=0.86]{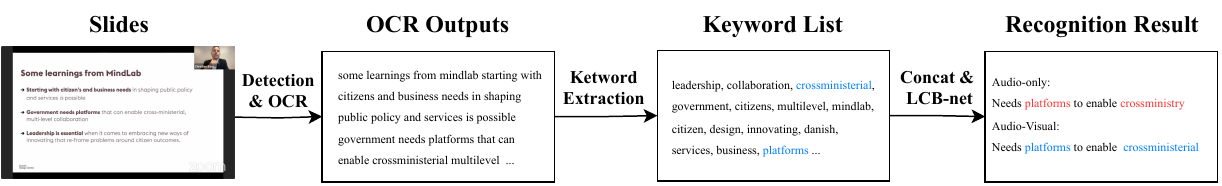}
        \vspace{-15pt}
	\caption{An overview of the AV-ASR system.
	}
	\label{pipline}
        \vspace{-15pt}
\end{figure*}

In contrast to regular contextual scenarios, the contextual phrases derived from slides within the visual stream (online conferences and courses) are synchronized in real-time with the speech, which implies inherent associations and strong contextual relationships among the phrases.
However, many existing methods utilize a contextual encoder to extract the bias embedding of each contextual phrase, which ignores the phrase relationships.
Therefore, in this paper, we propose a novel long-context biasing network (LCB-net) to effectively leverage the long-context information available in slide videos as the input of the context encoder.
Note that our context encoder employs token-level modeling and learns long-range dependencies among the contextual phrases.

Additionally, instead of computing cross-entropy (CE) loss for predicting the biased phrases that appear in the audio after ASR encoder (e.g, CPP~\cite{cpp2023Huang} and CIF ColDec~\cite{han2021cif}), we propose a biasing prediction network to explicitly determine biased phrases with binary cross entropy (BCE) loss after context encoder.
Our biasing prediction network allows the context encoder to effectively learn the relationships among biased phrases from long-context biasing.
Besides, we introduce a dynamic contextual phrases simulation during training, enhancing the diversity of the contextual phrases list.
Finally, we conduct a series of experiments over SlideSpeech~\cite{wang2023slidespeech} and LibriSpeech~\cite{panayotov2015librispeech} corpus to validate the performace of our proposed LCB-net.
The rest of this paper is organized as follows. Section \ref{sec:methods} describes our proposed model and methods. Section \ref{experiments} presents our experimental setup and results, while Section \ref{analysis} analyzes the performance of our model. The conclusions and future work will be given in Section \ref{conclude}.

\vspace{-5pt}
\section{PROPOSED METHOD}
\label{sec:methods}
\vspace{-3pt}
\begin{figure*}[!t]
	\centering
        \vspace{-6pt}
	\includegraphics[scale=0.5]{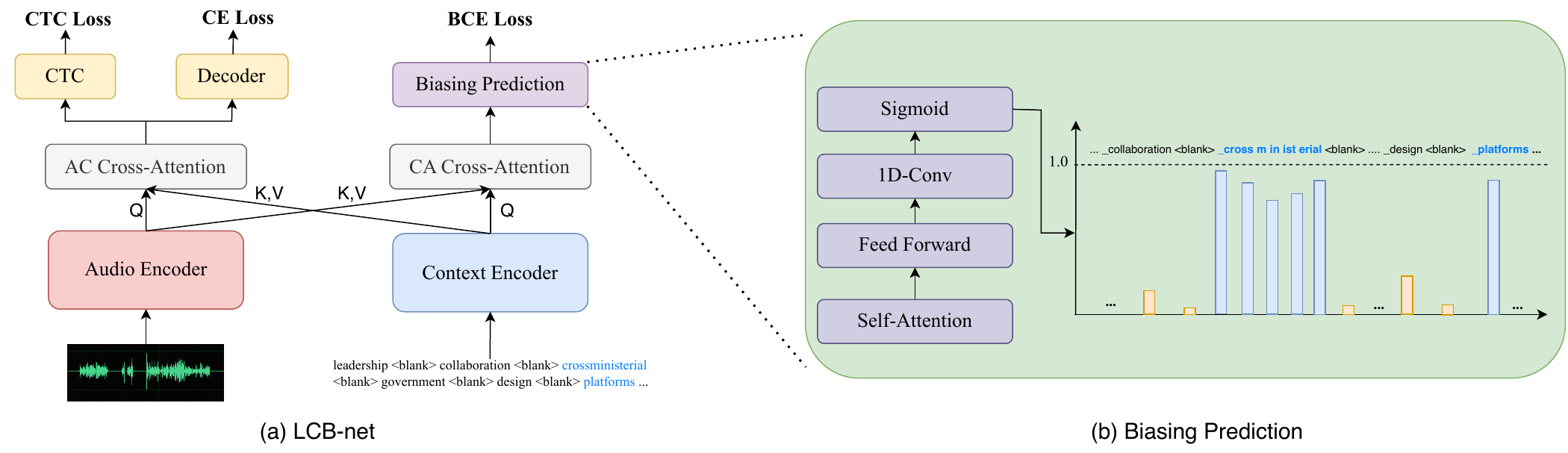}
        \vspace{-25pt}
	\caption{(a) LCB-net (b) Biasing Prediction
	}
	\label{model}
        \vspace{-14pt}
\end{figure*}

\vspace{-2pt}
\subsection{Overview}
\vspace{-3pt}

The overall framework of our text-based multi-modal AV-ASR systems is illustrated in Fig.~\ref{pipline}, which leverages video, text and audio information to transcribe the language content.
In detail, optical character recognition (OCR)~\cite{mithe2013optical} is firstly used to detect and recognize the text present in the slides. Next, we utilize keyword extraction methods~\cite{beliga2015overview} to capture the contextual biased phrases.
Lastly, we feed the both keyword list and audio into our LCB-net for recognition, which can effectively correct the biased phrases (the blue phrases in Fig.~\ref{pipline}) compared with the audio-only result.

\vspace{-4pt}
\subsection{Model Architecture}
\vspace{-3pt}
Our proposed LCB-net is depicted in Fig.~\ref{model}(a), which adopts audio and context bi-encoder architecture.
The LCB-net consists of seven modules, namely the audio encoder, context encoder, audio-context (AC) cross-attention, context-audio (CA) cross-attention, CTC, decoder and biasing prediction module.
Initially, the audio data ($\mathbf{X}=\{x_1,x_2,...,x_{n}\}$) and context data ($\mathbf{C}=\{c_1,c_2,...,c_{n}\}$) are processed through the audio and context encoder for representations learning (Eq.~(\ref{eq:encoder_asr},\ref{eq:encoder_context})).
To generate our long-context biasing data, we randomly shuffle the keyword list and concatenate the contextual phrases with \textit{$\langle blank\rangle$} tokens.
If there are no contextual phrases in the keyword list, we only add a \textit{$\langle blank\rangle$} token to the context encoder.
The AC cross-attention incorporates contextual biasing information from context encoder into the audio representations, which takes the audio representations as the query vector and the context encoder output as the key and value vectors (Eq.~(\ref{eq:ac_attention})).
On the other hand, the CA cross-attention takes the context encoder output as the query vector and the audio representations as the key and value vectors (Eq. (\ref{eq:ca_attention})), whose output feeds into the biasing prediction module to explicitly determine the biasing phrases.
Finally, we calculate CTC loss of AC cross-attention output and provide it to decoder to recognize tokens and calculate cross-entropy (CE) loss (Eq.~(\ref{eq:decoder})).

\vspace{-15pt}
\begin{align}
    & \mathbf{H}_{a} = \text{Encoder}_{a}(\mathbf{X}),  \label{eq:encoder_asr} \\ 
    & \mathbf{H}_{c} = \text{Encoder}_{c}(\mathbf{C}),  \label{eq:encoder_context} \\ 
    & \mathbf{H}_{ac} =  \text{CrossAttention}_{ac}(\mathbf{H}_{a}, \mathbf{H}_{c}, \mathbf{H}_{c}), \label{eq:ac_attention} \\
    & \mathbf{H}_{ca} =  \text{CrossAttention}_{ca}(\mathbf{H}_{c}, \mathbf{H}_{a}, \mathbf{H}_{a}), \label{eq:ca_attention} \\
    & \mathbf{Y} = \text{Decoder}(\mathbf{H}_{ac}). \label{eq:decoder}
\end{align}

\vspace{-16pt}
\subsection{Biasing Prediction Module}
\vspace{-2pt}

Fig.~\ref{model}(b) illustrates the structure of biasing prediction module, which consists of four parts: self-attention, feed-forward network, 1-dimensional (1-D) convolutions and sigmoid activation.
The biasing prediction module is designed to predict the biased phrases that appear in the audio.
Meanwhile, we hypothesize that both global and local interactions are significant for accurate prediction.
If the majority of BPE units within a word are determined as biased units, local modeling can effectively extract fine-grained local context patterns and determine all BPE units of the entire word as biased units (e.g. ‘\_cross m in ist erial’ is the BPE units of the word ‘cross-ministerial’, biased units ‘{\color{blue}\_cross m in} \# {\color{blue}erial}’ may help in determining the biased unit ‘{\color{blue}ist}’).
What's more, biased phrases also contain contextual relationships that can be enhanced through global modeling (e.g. ‘cross-ministerial’ may describe ‘platforms’, both may be biased phrases).
To incorporate these considerations, we first adopt self-attention (Eq.~(\ref{eq:self_atten})) and a feed-forward network (Eq.~(\ref{eq:ffn})) to establish long-range dependencies of the context representations, which have already fused acoustic knowledge from CA cross-attention (Eq. (\ref{eq:ca_attention})).
Then, we pass a window of $\mathbf{H}_{ffn}$ (e.g. $[ h_{l-2}, h_{l-1}, h_l, h_{l+1},h_{l+2}$]) to 1-D convolution for modeling local dependencies (Eq.~(\ref{eq:conv})).
The window size is set to 2, as most words do not exceed 5 BPE units.
Finally, we employ a fully connected layer with one output unit and a sigmoid activation to determine biased phrases with binary cross entropy (BCE) loss (Eq.~(\ref{eq:sigmoid})).
\vspace{-30pt}
\begin{align}
     & \mathbf{H}_{att} =  \text{SelfAttention}(\mathbf{H}_{ca} ), \label{eq:self_atten}\\
     & \mathbf{H}_{ffn} =  \text{FeedForward}(\mathbf{H}_{atten}), \label{eq:ffn} \\
     & \mathbf{H}_{cov} =  \text{Conv}(\mathbf{H}_{ffn}), \label{eq:conv} \\
     & \alpha  =  \text{Sigmoid}(\text{Linear}(\mathbf{H}_{cov})). \label{eq:sigmoid}
\end{align}
\vspace{-28pt}

\vspace{-5pt}
\subsection{Contextual Phrases Simulation}
\begin{figure}[!htb]
	\centering
        \vspace{-4pt}
	\includegraphics[scale=0.41]{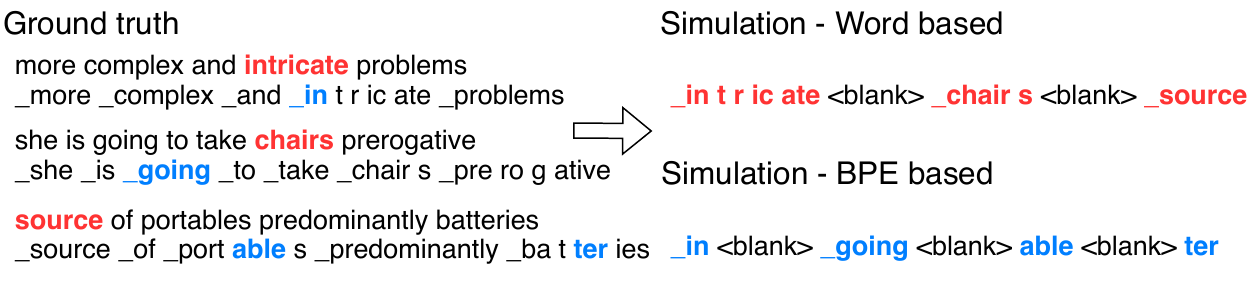}
        \vspace{-25pt}
	\caption{An overview of contextual phrases simulation.}
	\label{simu}
        \vspace{-15pt}
\end{figure}

\vspace{-2pt}
Contextual models with deep biasing, including our LCB-net, are all independent of various contextual phrase lists.
But in limited-scale data, the performance of contextual model is easily affected by the various contextual phrase lists, especially when the phrase numbers and lengths involved in the decoding and training period are different.
Word-based contextual phrases simulation~\cite{cpp2023Huang} was proposed to prevent the models from being overly dependent on limited contextual phrase lists.
Word-based simulation randomly selects entire contextual phrases from each utterance and combines them as the contextual phrases in the batch during training, as shown in Fig.~\ref{simu}.
Considering that our LCB-net employs the token-level modeling of BPE units, instead of using LSTM to merge each phrase into a single embedding (e.g. CPP~\cite{cpp2023Huang} and CLAS~\cite{deepcontext}), we propose a BPE-based simulation, which randomly selects a subset of BPE units from an entire word and and combines them as the biasing.
With these simulations, our LCB-net can easily generalize to variant contextual biasing phrases, leading to a more practical and stable solution.

\vspace{-4pt}
\section{EXPERIMENTS}
\label{experiments}
 \vspace{-4pt}
\subsection{Experimental Data}
\vspace{-3pt}
We use SlideSpeech corpus~\cite{wang2023slidespeech}, a \textbf{large scale} and \textbf{open-source}\footnote{Download at https://slidespeech.github.io/} slide-enriched audio-visual dataset, to evaluate the performance of our LCB-net.
The SlideSpeech corpus contains 1065.86 hours data for training (Train-L), 5.07 hours for validation (Dev), and 8.75 hours for test (Test).
To facilitate academic research, SlideSpeech corpus samples a subset (Train-S) from Train-L, comprising approximately 206 hours data.
The SlideSpeech corpus contains a significant amount of real-time synchronized slides with speech, even more than 94\% of the sections in test sets contain slides. 
To validate the model performance more fairly, we also conducted experiments on LibriSpeech corpus~\cite{panayotov2015librispeech}, which comprises 960 hours of English audiobook readings.
The dev-clean and dev-other sets were used for validation, while the test-clean and test-other sets were used for evaluation.

\vspace{-5pt}
\subsection{Experimental Setup}
\vspace{-1pt}

In all experiments, we use the 80-dimensional Mel-filterbank feature extracted with a 25 ms frame length and a 10 ms window shift.
For ASR and LCB-net models, audio encoder consists of 12-layer Conformer~\cite{gulati2020conformer}, and context encoder and decoder contains 12/6 Transformer~\cite{DBLP:booktitles/corr/VaswaniSPUJGKP17} layers.
The dimension of the attention and feed-forward layer is 256 with 4 heads and 2048, respectively.
SpecAugment~\cite{specaug} is applied for audio data augmentation and mask strategy~\cite{kenton2019bert} is applied for long-context biasing augmentation.
For both datasets, we first pre-train our LCB-net only with simulated phrase lists.
For SlideSpeech, we use the OCR results from slides along with simulated phrase lists for fine-tuning.
All models are trained on four 32GB memory V100 RTX GPUs, with a maximum trainable epoch of 75 and a warmup of 20,000 iterations.
The mix ratio of word and BPE based simulation strategies are set to 0.3 and 0.5, which are the optimal weights from the parameter tuning experiments.
During testing, we utilize the OCR results from slides as the SlideSpeech biasing lists and the \cite{le2021contextualized} provided lists as the LibriSpeech biasing lists.

We evaluate the results using WER, biased word error rate (B-WER) and unbiased word error rate (U-WER).
U-WER is measured on words that are \textbf{not in} the biasing list, while B-WER is measured on words that are \textbf{in} the biasing list~\cite{cpp2023Huang}.

\vspace{-12pt}
\subsection{Comparison to Different Contextual Models}
\vspace{-2pt}

Table~\ref{table1} shows the experiment results with or without contextual biasing lists of the contextual models trained on SlideSpeech Train-S.
Firstly, our proposed LCB-net outperforms general ASR model by 10.9\%/10.8\%/10.9\% and 9.4\%/9.1\%/10.9\% relative WER/U-WER/B-WER reduction on Dev and Test sets.
For the ablation study, we adopt the word-based simulation to improve the contextual learning, leading to 1.0\%/1.1\%/1.0\% and 1.5\%/2.0\%/0.8\% relative WER/U-WER/B-WER reduction on Dev and Test sets.
Meanwhile, BPE based simulation can further improve the generalization and robustness due to the BPE-level modeling of our LCB-net, which achieves the WER/U-WER/B-WER of 19.3\%/18.9\%/24.2\% on Test set.
When incorporating with the biasing prediction module, our LCB-net brings 2.4\% and 2.1\% relative B-WER reduction on Dev and Test sets.

Secondly, compared with CPP~\cite{cpp2023Huang}, which also uses the biasing simulation and fine-tuning strategies, our LCB-net with contextual biasing list leads to 9.6\%/10.4\%/2.4\% and 8.6\%/8.7\%/1.7\% relative WER/U-WER/B-WER reduction on Dev and Test sets.
Finally, without using contextual biasing lists, our LCB-net achieves relative 8.1\%/7.9\%/4.8\% and 7.0\%/6.7\%/4.1\% WER/U-WER/B-WER reduction on Dev and Test sets compared to CPP.
We believe that the context encoder in our LCB-net plays a crucial role in context knowledge transfer, enabling the ASR module to learn more language information from long-context biasing.

\begin{table}[!htb]
\centering
\vspace{-14pt}
\caption{Results of different contextual models on SlideSpeech Train-S. Reported metrics are in the following format: WER (U-WER/B-WER) (\%).}
\vspace{3pt}
\label{table1}
\scalebox{0.92}{
\begin{tabular}{lcc}
\toprule
\hline
\multicolumn{1}{c}{Model} & Dev                                                          & Test                                                         \\ \hline
ASR Baseline~\cite{wang2023slidespeech}              & 21.1 (20.3 / 31.3) & 21.2 (20.8 / 26.6) \\ \hline
\multicolumn{3}{l}{\textit{Inference with contextual biasing list (w B)}}   \\
\quad CPP w B~\cite{wang2023slidespeech,cpp2023Huang}                   & 20.8 (20.2 / 28.6) & 21.0 (20.7 / 24.1) \\ 
\quad LCB-net w B               & 19.6 (18.9 / 29.4) & 20.0 (19.7 / 24.7) \\ 
\quad \quad + Simu Word               & 19.4 (18.7 / 29.1) & 19.7 (19.3 / 24.5) \\ 
\quad \quad \quad + Simu BPE                 & 18.9 (18.2 / 28.6) & 19.3 (18.9 / 24.2) \\ 
\quad \quad \quad \quad + Bias. Pre.              & \textbf{18.8} (\textbf{18.1} / \textbf{27.9}) & \textbf{19.2} (\textbf{18.9} / \textbf{23.7}) \\ \hline
\multicolumn{3}{l}{\textit{Inference without contextual biasing list (w/o B)}}   \\
\quad CPP w/o B~\cite{wang2023slidespeech,cpp2023Huang}                 & 21.1 (20.3 / 31.4) & 21.3 (20.8 / 27.0) \\ 
\quad LCB-net w/o B             & \textbf{19.4} (\textbf{18.7} / \textbf{29.9}) & \textbf{19.8} (\textbf{19.4} / \textbf{25.9})\\ \hline
\bottomrule
\end{tabular}
}
\end{table}

\vspace{-25pt}
\subsection{Comparison to Different Training Data}
\vspace{-2pt}

As shown in Table~\ref{table2} and \ref{table3}, we trained our LCB-net on SlideSpeech Train-L and Librispeech, respectively.
The results demonstrate that our LCB-net outperforms general ASR model significantly on both datasets in WER/U-WER/B-WER.
For SlideSpeech Train-L, compared with CPP, our LCB-net brings 3.2\%/3.9\%/2.4\% relative WER/U-WER/B-WER reduction on Dev set.
For Librispeech, our LCB-net achieves 19.2\%/13.3\% relative U-WER reduction on two test sets compared with CPP, but there has been a performance decrease in B-WER.
We think that the difference between the two datasets is whether the biasing lists is highly correlated with the speech.
The biasing lists from SlideSpeech are synchronized in real-time with the speech, while that of Librispeech are simulated randomly. 
What's more, our LCB-net is stable and robust when using an empty biasing list.

\begin{table}[!htb]
\centering
\vspace{-8pt}
\caption{Results on SlideSpeech Train-L (\%).}
\vspace{3pt}
\label{table2}
\begin{tabular}{lcc}
\toprule
\hline
\multicolumn{1}{c}{Model} & Dev                & Test               \\ \hline
ASR Baseline~\cite{wang2023slidespeech}              & 13.1 (12.9 / 16.1) & 12.9 (12.9 / 12.7) \\
CPP w B~\cite{wang2023slidespeech,cpp2023Huang}              & 12.6 (12.7 / 12.4) & 12.4 (12.6 / 9.3)  \\
CPP w/o B~\cite{wang2023slidespeech,cpp2023Huang}                 & 12.9 (12.7 / 15.7) & 12.6 (12.6 / 12.6) \\
LCB-net w B               & \textbf{12.2} (\textbf{12.2} / \textbf{12.1}) & \textbf{12.0} (\textbf{12.2} / \textbf{9.0}) \\
LCB-net w/o B             & 12.6 (12.4 / 14.7) & 12.4 (12.4 / 11.3) \\ \hline
\bottomrule
\end{tabular}
\end{table}

\begin{table}[!htb]
\centering
\vspace{-12pt}
\caption{Results on Librispeech (\%).}
\vspace{3pt}
\label{table3}
\begin{tabular}{lcc}
\toprule
\hline
\multicolumn{1}{c}{Model} & Test-clean       & Test-other       \\ \hline
ASR Baseline~\cite{cpp2023Huang}              & 4.2 (2.6 / 18.1) & 8.9 (5.6 / 37.6) \\
CPP w B~\cite{cpp2023Huang}               & 3.4 (2.6 / \textbf{10.4}) & 7.8 (6.0 / \textbf{23.0}) \\
CPP w/o B~\cite{cpp2023Huang}                 & 4.3 (2.6 / 18.3) & 9.2 (5.9 / 37.5) \\
LCB-net w B               & \textbf{3.2} (\textbf{2.1} / 11.7) & \textbf{7.2} (\textbf{5.2} / 25.1) \\
LCB-net w/o B             & 3.7 (2.3 / 15.2) & 8.1 (5.4 / 32.0) \\ \hline
\bottomrule
\end{tabular}
\vspace{-12pt}
\end{table}
\vspace{-22pt}
\section{ANALYSIS}
\label{analysis}
\vspace{-5pt}

\begin{figure}[!htb]
\vspace{-6pt}
	\centering
	\includegraphics[scale=0.66]{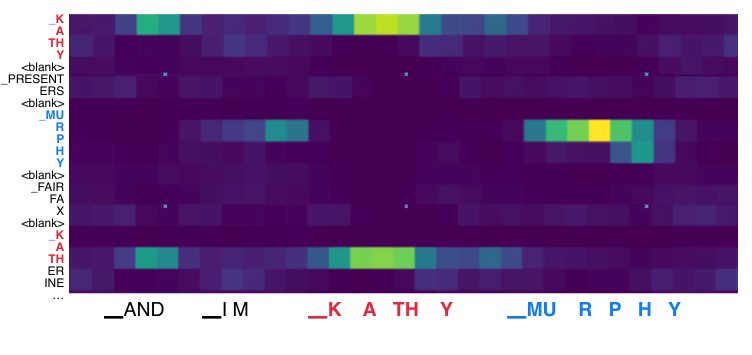}
	\vspace{-16pt}
	\caption{Attention score matrix of AC cross-attention. Brighter colors denote values closer to 1, while darker colors indicate values closer to 0. Red and blue mean the two biased phrases. Only the part  around the bright bias are plotted due to large bias number.
	}
	\label{attention_score}
\vspace{-0.2cm}
\end{figure}

To analyze the behavior of our proposed model, Figure~\ref{attention_score} visualizes the attention scores of our audio-context cross-attention.
We can observe that our LCB-net accurately predicts two biased phrases (personal names) and the biasing attention probability plays a critical role in this utterance.
Moreover, due to our fine-grained modeling units, other words (e.g. KATHERINE) with the same prefix (KATH) can also assist in predicting the biased phrase (KATHY).

\vspace{-16pt}
\section{CONCLUSION}
\label{conclude}
\vspace{-6pt}

In this paper, we propose a novel long-context biasing network (LCB-net) for audio-visual speech recognition (AVSR), which adopts a bi-encoder architecture to model audio and long-context biasing simultaneously.
Besides, we propose a biasing prediction module to explicitly determine biasing phrases in long-context biasing.
Experiments on SlideSpeech and Librispeech both prove that LCB-net outperforms general ASR model significantly.
Besides, we further analyze the performance with ablation study and attention score matrix.
In the future, we will investigate to train our model on industrial big data for real-world applications.


\begin{spacing}{0.96}
\bibliographystyle{IEEEbib}
\bibliography{strings,refs}
\end{spacing}

\end{document}